\documentclass[aps,prl,twocolumn,groupedaddress]{revtex4}
\usepackage{graphicx}
\usepackage{epsfig}
\usepackage{makeidx}
\usepackage{amsmath}
\usepackage{amssymb}
\usepackage{bbm,mathrsfs,nicefrac,units}
\newcommand{\s}{\mathcal{S}}
\newcommand{\buy}{\mathcal{B}}
\newcommand{\w}{\mathcal{W}}
\newcommand{\thr}{\mathcal{T}}
\newcommand{\pj}{\hat{p}_{\,\text{merge}}}
\newcommand{\pf}{\hat{p}_{\,\text{frg}}}
\newcommand{\ps}{\hat{p}_{\,\text{sell}}}
\newcommand{\pb}{\hat{p}_{\,\text{buy}}}
\newcommand{\sz}{s}

\begin{document}

\title{Computational modeling of collective human behavior: Example of financial markets} 
\author{Andy Kirou$^1$, 
B\l a\.{z}ej Ruszczycki$^1$, Markus Walser$^2$ and Neil F. Johnson$^{1,*}$}
\affiliation{$^1$Department of Physics, University of Miami, P.O. Box 248046, Coral Gables, FL 33124 USA\\
$^2$Landesbank Baden-W\"{u}rttemberg, Am Hauptbahnhof 2, 70173 Stuttgart, Germany\\
\\
{\rm {\bf Draft of keynote lecture at International Conference on Computational Science (June, 2008).}}\\
 {\rm {\bf  Final version published in LNCS M. Bubak et al. (Eds.) p. 33 (Springer-Verlag, Berlin, 2008)}}}

\begin{abstract}
We discuss how minimal financial market models can be constructed by bridging the gap between two existing, but incomplete, market models: a model in which a population of virtual traders make decisions based on common global information but lack local information from their social network, and a model in which the traders form a dynamically evolving social network but lack any decision-making based on global information. We show that a suitable combination of these two models -- in particular, a population of virtual traders with access to both global and local information -- produces results for the price return distribution which are closer to the reported stylized facts. We believe that this type of model can be applied across a wide range of systems in which collective human activity is observed.
\end{abstract}

\maketitle

\section{Introduction}
As a result of the increased availability of higher precision spatiotemporal datasets, coupled with the realization that most real-world human systems are complex, a new field of computational modeling has emerged in which the goal is to develop minimal models of collective human behavior which are consistent with the observed real-world dynamics in a wide range of systems. For example, in the field of finance, the fluctuations across a wide range of markets are known to exhibit certain generic statistical stylized facts  \cite{econo,bouchaud,stanley,jjh}. Socio-economic systems were traditionally treated from the points of view of game theory or traditional economic theory. These approaches -- while undoubtedly successful in terms of gaining insight into core features -- are unable to address the issue of how and why such systems produce the fluctuating external signals that they do \cite{econo,bouchaud,stanley,jjh}. 

Human systems as diverse as traffic, Internet downloads, and financial markets, are all known to produce
 large-scale fluctuations -- for example, in the number of cars taking a certain road, or the number of people accessing a certain website, or the number of people trying to sell a stock at certain times\cite{econo,bouchaud,stanley,jjh}. In previous decades, there was typically an insufficient amount of reliable data available for researchers to address such problems of dynamics. Nowadays, with the increase in online logging of data -- from social, governmental and commercial sectors -- this area of modeling now becomes very attractive. However few advances are likely to be made analytically, since any meaningful explanation of the dynamics must be related back to what the collection of individual objects are doing. In other words, it is what physicists call a many-body problem -- one in which the objects are subjected to endogenous and exogenous feedback and nonlinear interactions -- and it is known that such many-body problems are in general intractable. Given the additional feature that the objects themselves may be semi-autonomous (i.e. they each have some form of independent decision-making ability such that a given external input may yield various possible outputs depending on some internal state of the object itself) the most realistic route toward advancing our understanding of such systems is centered around computational modeling and simulation. 
 
The attraction of studying financial markets is that high frequency data can be obtained, albeit at a cost, and the underlying actions of any individual trader are ultimately quite simple: buy, sell or do nothing at any timestep. A wide range of interesting stylized facts have emerged based on analysis of the real market data over a wide range of timescales -- from seconds through to days, weeks and months. In particular, it has been found that the distribution of price returns (i.e. changes in price between a given time $t$ and a time $t+\Delta t$ later) do not follow the simple distribution expected from a random walk. The standard model of a financial market -- based on the efficient market hypothesis -- is that price-changes are like the toss of a coin. They are supposedly independent -- hence if we assume that each trader trades according to the toss of a coin, then the probability of buying and selling would a priori be $\frac{1}{2}$ if we ignore the `do nothing' option. Counting a head as $+1$ in terms of price change, and a tail as $-1$, the probability distribution for having a given price-change $\Delta P=N_{\rm buy}-N_{\rm sell}$ is simply the probability of  obtaining $N_{\rm buy}$ heads and $N_{\rm sell}$ tails from $N$ coin-tosses, such that $N=N_{\rm buy}+N_{\rm sell}$. This is a binomial expression, which then approaches a Gaussian in the large-$N$ limit. 

As a rough first approximation, financial market behavior is not far from the Gaussian model. However, many independent detailed empirical studies of financial market returns have confirmed that major deviations begin to arise in the tails of the distribution\cite{econo,bouchaud,stanley,jjh}. Specifically, the distribution of price-changes $\Delta P$ deviates from Gaussian behavior at moderate values of $\Delta P$. In particular, the probability of intermediate-to-large price-changes is larger than the random coin-toss model would suggest. This leads to the `fat tail' description often aimed at financial markets -- such fat-tailed behavior is also  common across a wide range of socio-economic domains \cite{econo}. The fact that large price-changes are more likely than expected -- and in more general socio-economic settings, that large traffic jams or heavy Internet downloads are more likely than expected --  suggests that the population is unintentionally behaving in a coordinated way. It is as though the supposedly independent coin-tosses of the $N$ traders are not in fact independent: when one comes up heads, they are all more likely to come up heads, and vice versa with tails. The fact that getting the best price in a financial market is a competitive activity -- in the same way that managing to grab space on a busy road, or a download on a busy website, are also competitive -- means that such coordination is very unlikely to have arisen through some intentional population-level decision making. There is no central controller -- and even if there were, the fact that individual objects are competing to win means that no central controller would necessarily be listened to or followed. This coordination observed in many scenarios where populations of humans are competing for some limited resource, thereby leading to larger than random probability for large events, is characteristic of many human systems. But if so, what causes it and how can we provide a quantitative model of it? Because of the generic nature of this fat tail statistics, any model cannot depend on the details of the particular market or type of trader, or road or type of car, or website or type of computer. Instead it must be some fairly general feature of collective human activity. 

The fundamental question as to what underlying model might best represent such collective coordination, has inspired a new breed of computer-based scientific investigation involving physical, biological and social scientists. At its root, a system such as a financial market, traffic system, or the Internet, involves agents (i.e. people) deciding between a few options (e.g. buy, sell, do nothing) based on some limited information -- which may be global or locally generated -- and then competing with the remaining agents for the available resource or reward. Collective coordination requires some form of trader inter-connectedness. One way in which this could have arisen, is if subsets of the traders form social groups such that they and their immediate friends or associates, coordinate their actions (i.e. bias their decisions and hence effectively connect their coins during the coin-toss). In fast-moving financial markets, such groups are likely to change fairly rapidly, and should at least be accounted for using dynamical models of such group formation. This idea has led to a particular class of models based on dynamical cluster formation. Notable examples include  
the dynamical clustering model of Egu\'iluz and Zimmermann\cite{ez} and of others in biology \cite{gl}\cite{cb}. A second way in which coordination could have arisen, concerns how agents react to a particular piece of common information. Models of this form,  of which a notable example is the so-called
Grand Canonical Minority Game \cite{gcmg}, feature agents whose actions are dictated by a strategy (or set of strategies). In the first case of real clusters, the grouping is intentional, while in the second case agents form unintentional groups (i.e. crowds) as a result of using the same strategy at the same time and hence acting identically over a short period of time.
These two classes of model are complementary. Each of these models is `minimal' in the sense that they are the simplest known examples which seem to capture the essential ingredients of clustering and decision-making respectively. To date, the two classes of model have been studied separately -- however, they should clearly both be combined in order to understand the interplay of local and global information on collective group formation and hence the collective dynamics. 

In this paper, we analyze a collective human system in which there are simultaneously {\em local} interactions as observed in the Egu\'iluz and Zimmermann\cite{ez} model (i.e. E-Z model) and {\em global} interactions as observed in the Grand Canonical Minority Game \cite{gcmg} (i.e. GCMG). 
The focus of this paper is on how adding global interactions to the E-Z model, does indeed improve the fit with the known empirical distributions of financial market returns.
Our construction is a modified E-Z model in which the agents are randomly assigned the trading strategies and apply their strategy based on the last two price movements.
Since there is no strategy score introduced yet, the system so far does not have
effectively a memory and behaves as relative majority vote system. 
\begin{figure*}[t]
\centering
\epsfig{file=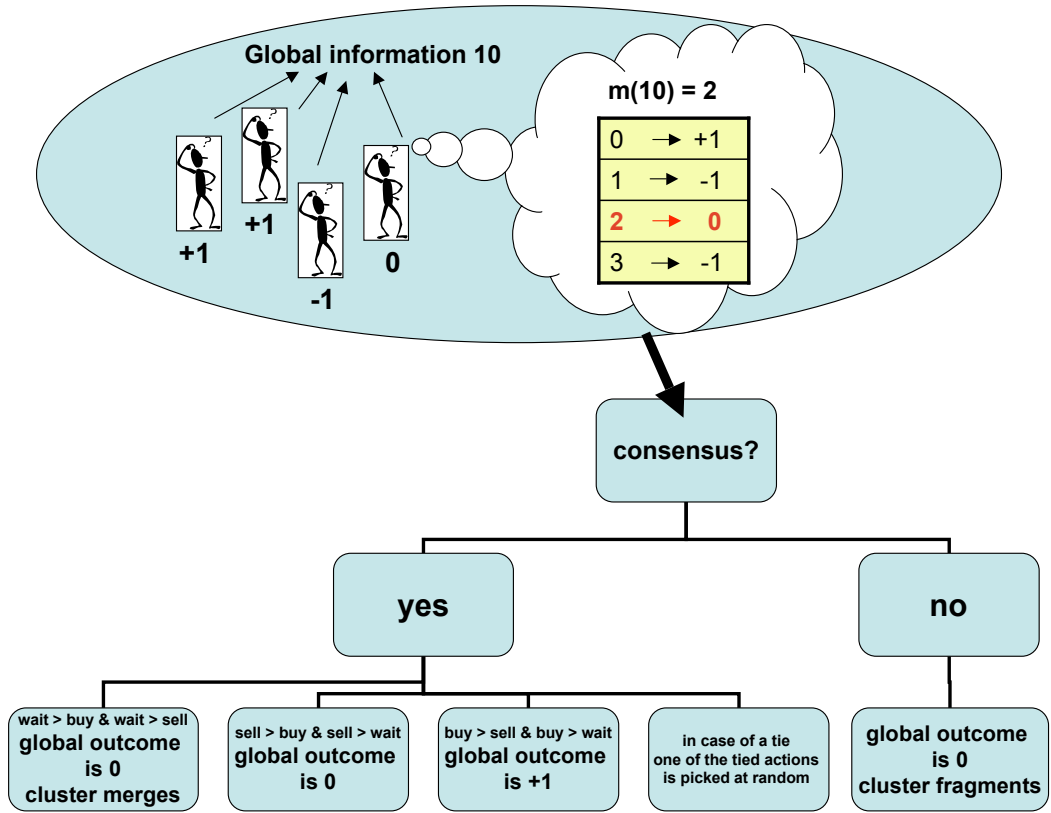,width=0.6\textwidth,angle=90}
\caption{Schematic diagram of the decision process.}\label{fig:1}
\end{figure*} 
 
\section{Relative Majority Vote}  
    
The relative majority vote system is specified by two parameters, the total number of agents $N$ and
the consensus parameter $x$.
At each time-step (to set up a time scale we need to prescribe some time-scale constant)
an agent is chosen at random and the group to which
the agent belongs to is pointed. The size of this group is denoted as $\sz$.
Once a particular group has been pointed to it makes a decision in the following way: Each agent votes either to
\textit{sell}, \textit{buy} or \textit{wait} (see Fig. 1) with approximate probability $\nicefrac{1}{3}$ for each of the listed options.
If the number of votes for the most popular decision exceeds the
threshold, which is defined as 
\begin{equation}
\thr\equiv x \cdot \sz
\end{equation}
the consensus is reached and the decision is performed, otherwise there is no consensus and
the group breaks into single individuals. The group that decides to wait does not trade but joins another group
by choosing randomly any other agent from the entire population and merging with a group to which this agent belongs.
Let us for a particular group of size $\sz$ 
denote by $\s,\buy,\w$ the number of agents who vote to sell, buy or wait (and merge with another group) respectively.
The conditions for the group decision are precisely      
\begin{equation}
\begin{array}{lllll}
\text{Fragments:}&\quad &(\w<\thr)&\wedge\, (\s<\thr) &\wedge \;(\buy< \thr)\\
\text{Buys:}&\quad &(\buy \geq \thr)&\wedge\,(\buy > \s) &\wedge \;(\buy > \w) \\
\text{Sells:}&\quad &(\s \geq \thr)&\wedge\,(\s > \buy) &\wedge \;(\s > \w) \\
\text{Merges:}&\quad &(\w \geq \thr)&\wedge\,(\w > \buy) &\wedge \;(\w > \s) 
\end{array}\label{rules}
\end{equation}
We need also to account for the fact we may have tied number of votes, which is resolved by randomly picking one 
of the "tied" decisions, e.g. if
\begin{equation*}
(\s\geq\thr)\wedge(\s=\buy)\wedge(\s>\w)
\end{equation*}
then the group either sells or buys.
The decision presented in \eqref{rules} are exclusive, 
therefore for the corresponding conditional probabilities (on the condition that the particular group is chosen)
it holds that 
\begin{equation}
\pf+\ps+\pb+\pj=1\,,\label{complete}
\end{equation}
where $\pf$ is the probability that the group fragments.
Note that the above conditional probabilities depend on the group size $\sz$.
From the symmetry of \eqref{rules} and \eqref{complete} we see that is sufficient to know $\pf$ since
\begin{equation}
\ps=\pb=\pj=\frac{1-\pf}{3}\,.\label{probs}
\end{equation}
We calculate the combinatorial expression as
\begin{equation}
\pf(\sz)=\frac{\sz!}{3^\sz}\sum_{\w=0}^{\sz-1}\;\sum_{\buy=\sz-\thr-\w}^{\min(\thr-1,\sz-\w)}\frac{1}{\w!\buy!(\sz-\buy-\w)!}
\label{pf}
\end{equation}
\begin{figure*}[t]
\centering
\epsfig{file=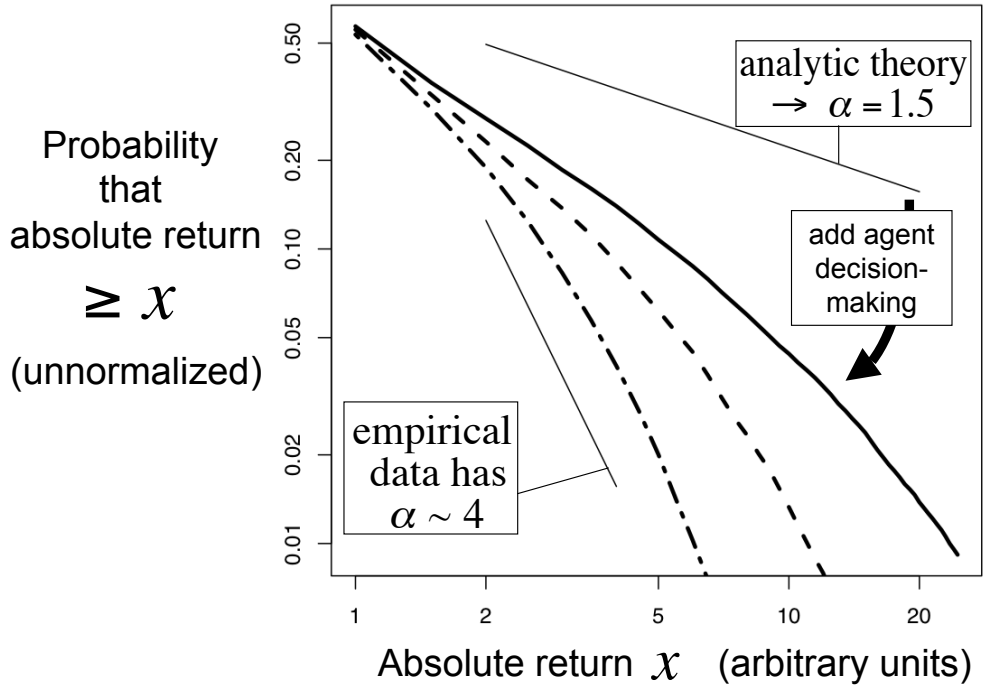,width=0.6\textwidth,angle=90}
\caption{Cumulative return distribution in arbitrary units for consensus parameter \hbox{$x=37\%$ (solid line)}, 
\hbox{$x=41\%$ (dashed line)} and \hbox{$x=47\%$ (dashed-dotted line)}. The number of agents used was 10000. The power-law exponents refer to the underlying return distribution, and hence differ from the cumulative distribution value by unity}\label{fig:2}
\end{figure*}
The system may be described by the mean field theory, disregarding the fluctuations and finite size effects. 
We denote by $n_{\sz}$ the average number of groups of size $\sz$. For the steady state the set of master
equations is semi-recursive\footnote{The equation for $n_l$ depends on $n_i$ for $i=1\ldots l-1$ and 
on a constant being a function of all
$n_i$'s} and is written as
\begin{align}
-&\frac{\sz}{N}\pf(\sz) (1-\delta_{\sz 1})n_{\sz}\notag\\
-&\left[\frac{1}{N}\pj(\sz)
+
\frac{1}{N^2}\sum_{\sz '=1}^{N}\sz' \pj(\sz')\right]\sz\,n_{\sz}\notag\\
+&\sum_{\sz'=1}^{\sz-1} \,\sz'\,n_{\sz'}\,(\sz-\sz')n_{\sz'}\pj{\sz'}\notag\\ 
+&
\delta_{\sz,1}\sum_{\sz'=\sz+1}^N  \pf(\sz')\sz'^2n_{\sz'}=0\,.\label{meq}
\end{align}
The above set may be solved numerically \footnote{The numerical procedure of solving~\ref{meq} is effective
for limited values of $N$ as at least $\nicefrac{1}{2}\,N^2$ iteration steps are required}. To be able to model
not just the mean behaviour but also the corrections coming from the fluctuations, we need to perform direct simulations 
of the system described by \eqref{meq}.

We are interested in the situation when the consensus parameter is within the range of  
$x=(33\,\nicefrac{1}{3}\;\%,50\,\%)$. 
If $x<33\,\nicefrac{1}{3}\;\%$ there is no fragmentation,  if $x>50\,\%$ we have the absolute majority vote condition. When 
$x\rightarrow 33\,\nicefrac{1}{3}\;\%$ from above we expect that the return distribution approaches the one for the original 
Egu\'iluz-Zimmermann system, which shows a power law with exponent $1.5$ over large scale of return sizes.
As we increase $x$ slightly over $33\,\nicefrac{1}{3}\;\%$ the dominant behavior of the buy/sell probability $\pb=\ps$ for the large
groups (of order $\sz\succapprox 100$) is an exponential cut-off, while for smaller $\sz$ we have the finite size effect.
This modifies the model in two ways: The group distribution changes since the conditional probabilities enter Eq. \eqref{meq}, 
and the trade mechanism changes. Unlike the E-Z system \cite{ez} (where mostly the large groups trade) we expect to 
have the trades coming from the actions of the small groups with the exponential cut-off due to the behavior of Eq. \eqref{pf}.     

\section{Simulation Results}
The simulations were performed for a system with $N = 10^5$ agents, $m = 2$, and $10^6$ time steps, with 
three different values of the consensus parameter. The initial state of information was $(1,1)$. 
After $10^5$ time steps (in order to allow the system to reach equilibrium) the returns where computed as follows: 
If a cluster of size $\sz$ decided to buy, the return was $+\sz$. If a cluster of size $\sz$ decided to sell, the return 
was $-\sz$. After the simulation was complete, the time was rescaled by adding the returns of two consecutive time steps 
since on average a transaction occurred once every two time steps. Thus, the results in Fig.~\ref{fig:2} are for 
$9*10^5/2$ time steps effectively.
It is observed that indeed most of the trades come from the action of the small groups. While 
the consensus parameter is increased the distribution of the returns is falling more sharply due to the dominance 
of the exponential cut-off.
The results are shown in Fig.~\ref{fig:2}. These results demonstrate that the addition of access to global information and subsequent decision-making into a model built primarily around local group formation, leads to a hybrid model which can better capture features of the known empirical distributions. In short, both local group formation {\em and} global information are important when building a minimal computational model of financial markets -- and, by extension, collective human activity in any domain in which competition exists between a collection of interconnected agents.

\section{Discussion}
We have explored a simple system featuring the combination of locally and globally interacting agent-based
models, as part of a more general quest for minimal computational models of real socio-economic systems based on individual-based behavior. Such minimal models aim to incorporate the minimum number of features (and hence parameters) that make the individuals' behavior and interactions appear credible, and yet are consistent with the maximum number of empirical stylized facts based on real-world data. 
In particular, we have proposed a simple construction which features
global interactions (via the heterogeneity of strategies held by the agents)
as well as agent memory in the locally interacting system (the grouping model). 

Our specific results are as follows. The scenario in which the agents are allowed to vote introduces the exponential cut-off starting at the scale where the
effects connected to the discrete nature of the system may be neglected. Our results show that those who usually trade are 
the small groups, and that there are no trades coming from the large groups.
By contrast in the original E-Z model, the conditional probabilities are constant and any particular 
large group trades more often than a particular small group. Since
the group distribution shows a power law behavior over a large scale of group size, 
the action of the large groups occurs more often than is indicated by empirical data. The most realistic minimal model (which is as yet undiscovered) should lie somewhere in between.
Any voting scenario is a Poisson process which introduces the exponential cut-off to the system appearing at a scale 
where the number of individuals involved is large enough that we may disregard the discrete nature of the system.
The modeling challenge is therefore to make the
decision-making process reflect more complicated behavior of the individuals such as possession of memory, behavior based on past experience,
and passing the information between groups about whether to trade or not. 

In terms of more general issues of computational modeling, we have tried to highlight the need to develop minimal computational models of real socio-economic systems through individual-based behavior. Future theoretical developments in such fields lie beyond simply integrating some form of phenomenological equation. Moreover, this sort of socio-economic modeling is an application of computation that is set to boom in the future given the growing availability of high-frequency data from socio-economic systems -- and the fundamental philosophical need for theories which treat {\em dynamical fluctuations} in addition to mean behavior. One particular example in which this philosophy is now being developed, is in improving our understanding of human conflict -- by looking at the stylized facts of conflict dynamics in exactly the same way as has been done for financial markets. Indeed, we have recently shown that remarkably similar minimal computational models can be built, with equally satisfying agreement with empirical data, simply by combining together global and local interactions among agents. This work on human conflict will be discussed in more detail elsewhere. 

\section{Acknowledgement}
NFJ is very grateful to Pak Ming Hui (Chinese University of Hong Kong) for discussions.

\end{document}